# Improving Light Yield Measurements for Low-Yield Scintillators

J.B. Cumming,[1,*] S. Hans,[1,2] and M. Yeh[1]

[1]*Chemistry Department, Brookhaven National Laboratory, Upton, New York 11973*

[2]*Chemistry Department, Bronx Community College, Bronx, New York 10453*

Light *power* spectra are introduced as a new tool for relative light yield (LY) determinations. Light *event* spectra have commonly been used for this purpose. Theoretical background supporting this change is provided. It is shown that the derivative of a light power spectrum can provide a reliable LY measurement at levels as low as 2% of those for high-yield liquid scintillators. Applications to light evolution in the PPO+LAB system and to water-based liquid scintillators are described.

Keywords: liquid scintillator, light yield, data analysis, water-based scintillator

## I. INTRODUCTION

Light yield (LY) measurements played an important role in developing liquid scintillator (LS) for the Daya Bay, RENO, and Double CHOOZ [1–3] reactor antineutrino experiments, both the conventional LS and that with added Gd (Gd-LS). They confirmed the long term stability of both scintillators before deployment, and provided quality control during the production of the multi-ton quantities required for the working detectors.[4, 5] A generic liquid scintillator consists of three components: a solvent, a primary flour, and a secondary flour (or "wavelength shifter"). The solvent is the first member of the chain that converts ionization produced by a particle into light. Presence of aromatic rings in the solvent, i.e., conjugated systems with delocalized $\pi$ electrons, is essential for a high LY organic liquid scintillator. Light from the solvent is typically too blue for good overlap with the sensitivity curve of a photomultiplier tube (PMT) . A frequently used primary flour is 2,5-diphenyl-oxazole (PPO). This absorbs the solvent light and emits it red shifted. The wavelength shifter, commonly 1,4-bis-(o-methylstyryl)-benzene (MSB), makes an additional upward shift for a still better match to the PMT response. Linear alkyl benzene (LAB) has been selected by several experiments as their solvent on the basis of its favorable optical properties, and low toxicity, flammability, and corrosiveness. LAB is an intermediate in detergent manufacture. A typical component is dodecyl-benzene, $C_{18}H_{30}$.

Daya Bay studies the properties of reactor-generated electron antineutrinos, $\overline{\nu}_e$, via the reaction: $\overline{\nu}_e + p = e^+ + n$. Gadolinium is added to the LS to detect the product neutrons. Hydrogen in LAB serves both as a target for the neutrinos and to thermalize the neutrons. A delayed coincidence between an $e^+$ and the capture $\gamma$-ray flash is a robust trigger for a $\overline{\nu}_e$ interaction. Light yield measurements played an important role in developing this metal loading. See Buck and Yeh [7] for a review of liquid scintillators and metal loading, .

Compton scattering of $\gamma$ rays has long been used to excite samples for LY measurements, e.g., Kallmann and Furst[8–10] have studied a variety of solvent-fluor systems. Light yields are routinely measured at Brookhaven using this method. The emitted light is collected, digitized, and stored. The half-peak-height point on the upper edge of a spectrum is taken as a marker of the Compton edge[11]. Safari et al.[12] have recently proposed that the LY derivative is a superior marker. For a Gaussian edge (mean=$m$, standard deviation=$\sigma$) , the 50% point occurs at $m + 1.17\sigma$, while the derivative maximum [1] is at $m + \sigma$. A derivative peak may be observed even if there is no resolved Compton peak, see below.

This paper introduces a new approach to relative light yield measurements, a "power derivative" method. This entails a simple transformation of an observed light event spectrum into a power spectrum. The derivative maximum then provides the Compton-edge marker. Background for this procedure is given in Section II. The use/performance of a commercial instrument for routine LY measurements is described in Section III. Section IV-A provides examples for high- and low-light-yield materials and for instrument blanks. The dependence of LY on PPO concentration is examined in Section IV-B. Over the range 0.1 to 10 g/L, data are consistent with a two exponential model: Strong absorption of LAB light by PPO is followed by weaker absorption of the resulting PPO light (self-quenching). Applications to water-based liquid scintillators (WbLS) are presented in Section IV-C.

## II. BACKGROUND

$^{137}Cs$, 30-year half life, is a convenient $\gamma$-ray source. A 661.7-keV $\gamma$ ray follows 85.1% of the $^{137}Cs$ beta decays. The interaction of 661.7-keV $\gamma$ rays with low Z

*Corresponding author: Cummingd@bnl.gov

[1] To facilitate comparison with peaks in the event and power spectra, the sign of the derivative will be reversed. When we refer to a derivative "maximum " or "peak", it is understood that the actual derivative is negative and at a minimum.

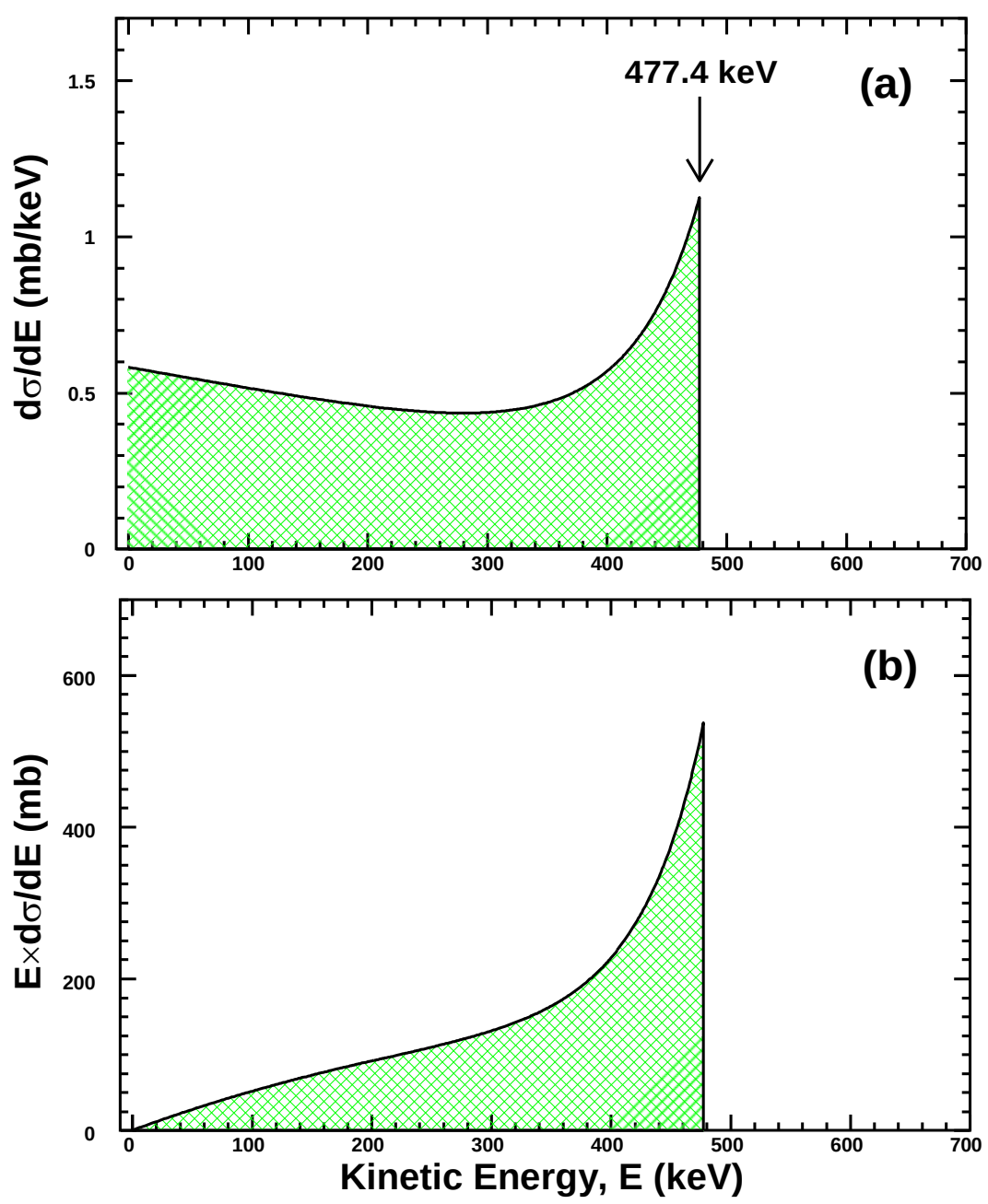

FIG. 1: Compton electron spectra for the scattering of 661.7 keV $\gamma$ rays by free electrons (a): Differential cross section as a function of energy. (b): Energy deposited (Power) as a function of energy.

elements is dominated by Compton scattering. Klein and Nishina [13] have analyzed the scattering of $\gamma$ rays by free electrons in the framework of quantum electrodynamics. Transforming their result to electron energy and integrating over all space yields the familiar Compton electron spectrum, $d\sigma/dE$ vs $E$, shown in Fig. 1(a). The sharp cutoff at 477.4 keV, the Compton edge, corresponds to backscattering of an incident $\gamma$ ray. The shaded area gives the total scattering cross section. Multiplication by the $\gamma$-ray flux and the electron density of the sample converts $d\sigma/dE$ to an electron current. Fig. 1(a) becomes a current vs voltage plot. The area under the curve becomes the total current through the sample. Further integration over the data acquisition time gives the total number of electrons, $N_{tot}$, produced in the sample during that time. We will refer to a distribution such as Fig. 1(a) as an "Event" spectrum as its area is a number. Note that $N_{tot}$ is proportional to the electron density not the matter density of a sample. Since Z/A=1 for H and Z/A=1/2 for C, more interactions per gram are expected for higher H-content materials.

Compton electrons lose energy by ionization of the solvent. The estar program[14] provides energy loss, $dE/dx$, and range, $R$, for a variety of materials. $dE/dx$ depends on the electron energy, the mean ionization potential and the density of the medium, e.g., $dE/dx$ for LAB is 3.2% greater than that for $H_2O$ at 477.4 keV. Since $R = \int_0^E (dE/dx)^{-1} dE$, all an electron's energy is ultimately deposited in the medium. The range of a 477.4-keV electron in LAB is 160 mg/cm$^2$, or 1.9 mm, comparable to the 24-mm inside diameter of our sample vials. Loss of energetic electrons from the sample and addition of low-energy electrons from the vial will distort the ideal electron spectrum. That the target electrons are not free is another source of distortion[12].

The electron event spectrum, Fig. 1(a), does not give the energy deposited in the sample. By electrical analogy, the energy deposited (=power) is the product of current times voltage. The energy deposition spectrum, $E\times(d\sigma/dE)$ is shown in Fig. 1(b). We will refer to this as a "Power" spectrum. Its integral gives the total power, $P_{tot}$, deposited in the sample. As expected, this transformation does not shift the Compton edge, but the peak is sharpened and the contribution from lower-energy electrons is much suppressed.

The novel contribution of the present work is to apply this same transformation to experimental LY spectra. Consider a LY determination as analogous to an efficiency measurement for an electrical transformer. Nonradiative transitions and light losses in the scintillator are equivalent to thermal losses in the transformer. When measuring the efficiency of a transformer, the relevant parameter is not the output-to-input voltage or current ratio, but the power ratio. We will show that the use of power spectra enables routine LY measurements on small samples in a laboratory environment at levels not accessible using event spectra.

## III. EXPERIMENTAL

A Beckmann LS-6500 Scintillation Spectrometer was used for our light yield measurements. This device is intended for assaying the $\beta$-activity of isotopes such as $^3$H, $^{14}$C, and $^{32}$P. In that application, an aliquot of the unknown is added to "scintillator cocktail" in a sample vial. When in the assay position of the LS-6500, the vial is viewed by two photomultiplier tubes. A prompt ($<$ 20 nsec) coincidence between the PMTs is required to trigger digitization and storage of their sum signal. This "two photon" trigger reduces a variety of background effects. Its influence on LY measurements is examined below.

Processing samples for beta assay occasionally produces color or material which reduces light output from the scintillation cocktail (color or chemical quenching). The LS-6500 can be operated in a "quench mode" in which an internal $^{137}$Cs source is moved from a shielded location to a position near the sample. Comparison of the light spectrum from the sample with that of an unquenched reference provides a measure of the quenching. This quench mode is essentially a LY measurement. To avoid irreproducibility in positioning, the movable source was replaced with a fixed one.

Samples for assay, 8.6 g by weight, were placed into standard 20-ml screw-cap Pyrex scintillation vials. This corresponds to 10 ml for LAB ($\rho$ = 0.86 g/cm$^3$). The LS-6500 can be programmed to analyze a set of vials sequentially, with the sequence repeated as desired. A data acquisition time of 30 min/sample served to balance throughput with statistics.

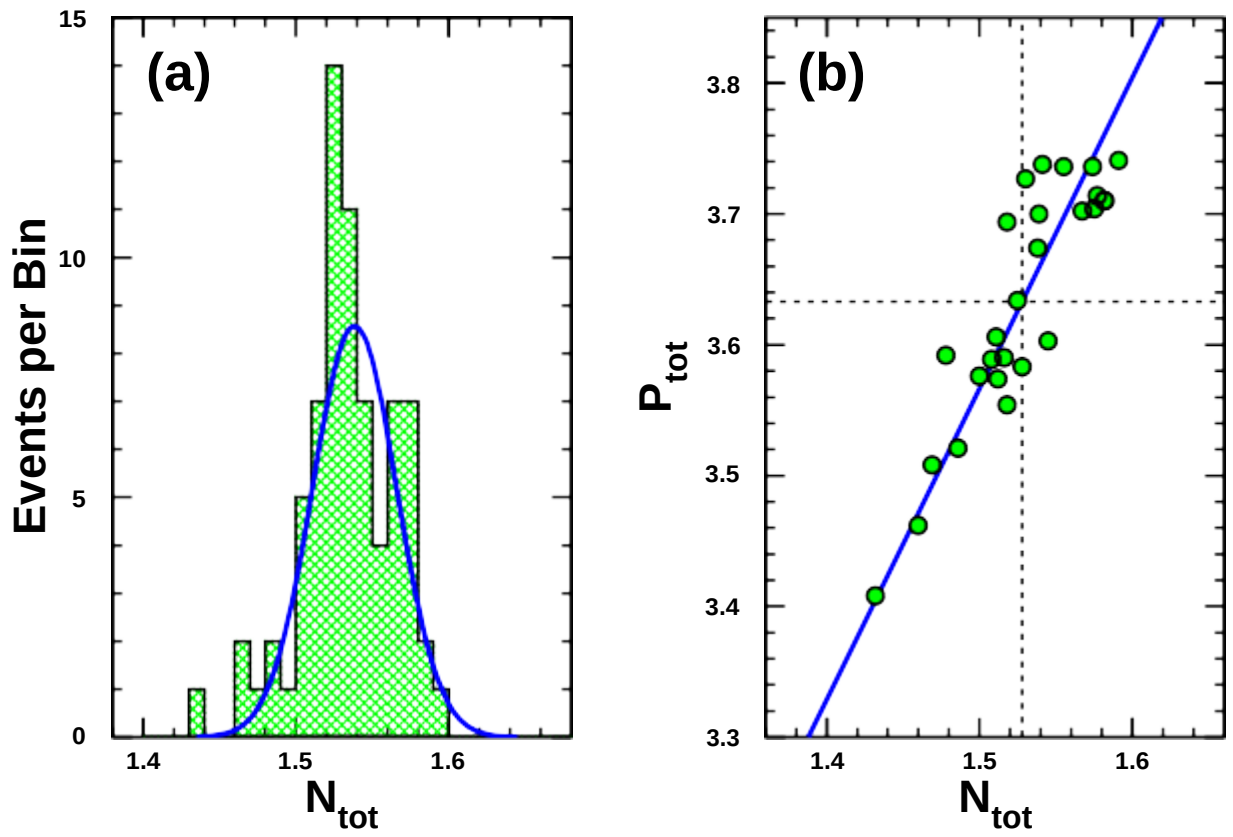

FIG. 2: (a): Histogram of observed $N_{tot}$ for 72 spectra expected to have identical values. The solid curve shows a Gaussian fit. (b): Correlation between $P_{tot}$ and $N_{tot}$ for 27 LY spectra of LAB containing 2.5-g/L PPO. Dashed horizontal and vertical lines indicate means of the distributions. A solid line shows the correlation expected if $\Delta P_{tot}$ were proportional to $\Delta N_{tot}$. Pearson's correlation coefficient is 0.89.

More than 200 LY spectra were acquired over an 9-month period. Many of these were replicates, permitting estimates of overall reproducibility of the measurements. Figure 2(a) is a histogram for 72 LY event spectra which were expected to have the same $N_{tot}$. The distribution is consistent ($\chi^2/ndf$ = 1.1) with a Gaussian having a mean $m = 1.538 \times 10^6$ and a $\sigma/m$ of 1.8%. This value is much larger than the < 0.1% expected from the number of events.

The correlation between $P_{tot}$ and $E_{tot}$ for a subset of 27 nominally identical spectra is examined in Fig. 2(b). Once again, deviations from the means are much larger than expected from statistics. The solid line shows the variation expected if $\Delta P_{tot}$ were proportional to $\Delta N_{tot}$. Pearson's linear correlation coefficient is $r$ = 0.89. A probable cause of these correlated errors is variation in the vial position relative to the $^{137}$Cs source. Uncertainty in derivative positions is significantly lower. An analysis of 183 power spectra gave $\sigma/m = 0.7\%$ for the derivative position.

## IV. RESULTS

### A. High and low light yields, blanks

A light yield event spectrum for typical liquid scintillator (3-g/L PPO, 15-mg/L MSB in LAB) is shown in Fig. 3(a), the derived power spectrum in Fig. 3(b). A trivial blank has been subtracted, see below.

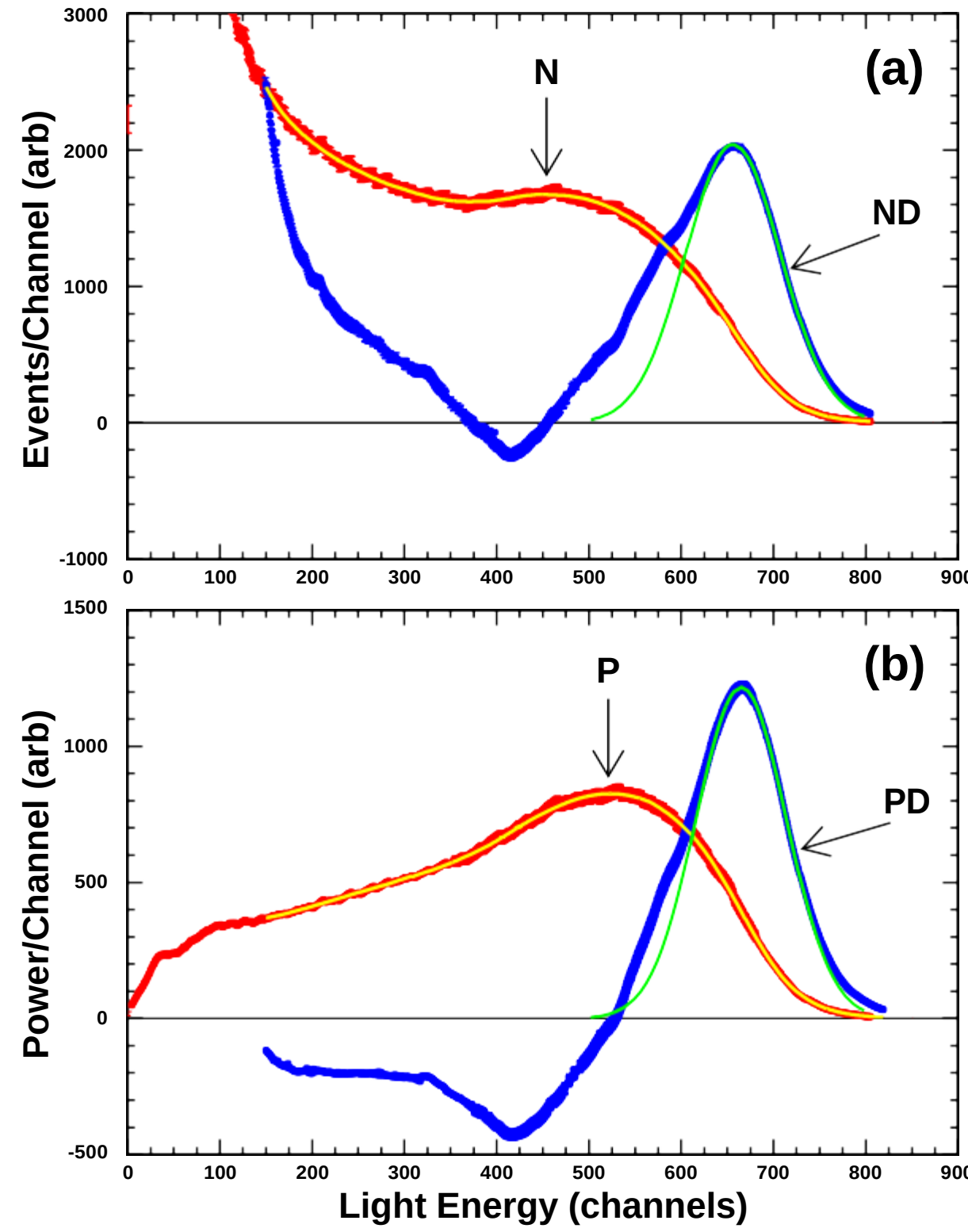

FIG. 3: Light yield spectra for a typical liquid scintillator; the event spectrum (N)) and its derivative (ND) in (a), and the power spectrum (P) with derivative (PD) in (b). See text for additional details.

Although the LS event spectrum contains $1.54 \times 10^6$ events, most are in a low-energy peak, off-scale by a factor of > 3 in Fig. 3(a). (Peaks at channels 3-5 are seen in all net event spectra obtained with the LS-6500 system.) Statistics at the upper end of the spectrum are limited.

The event and power spectra, N and P in Fig. 3, were smoothed using the function $\ln y = p_1 + p_2 x + p_3 x^2 + p_4 x^3$. The resulting curves are shown as a light lines superimposed on the observed points. The smoothing procedure also provided the derivative spectra, ND and PD, shown in the figure. Channel maxima, $ND_{max}$ and $PD_{max}$, were obtained from Gaussian fits to the top 20% of each derivative spectrum. The full Gaussians are shown as a light lines.

Comparison of Fig. 3 with Fig. 1 reveals the drastic effects of system resolution. Sharp Compton edges are transformed to broad Compton peaks. The Compton peak is barely resolved from the prominent low-energy peak in Fig. 3(a). Derivative maxima coincide with the Compton edges in Fig. 1. They are well above the Compton peaks in Fig. 3, where they mark inflection points on broad high-side falloffs.

To examine the effect of aliphatic chain length on light yield, LAB in the LS was replaced by toluene ($C_7H_8$).

TABLE I: Comparison of procedures for calculating the light yield (LY) of a scintillator with toluene ($C_7H_8$) as the solvent, relative to one based on LAB. Each contained 3 g/L of PPO and 15 mg/L MSB.

| Event Spectra | LAB based | $C_7H_8$ based |
|---|---|---|
| Peak channel | 454 | 521 |
| Zero crossing | 455 | 525 |
| Derivative peak | 656 | 719 |
| 50% Max at | 640 | 709 |
| **Power Spectra** | **LAB based** | **$C_7H_8$ based** |
| Total, $P_{tot}$ | $3.72 \times 10^8$ | $4.07 \times 10^8$ |
| Peak channel | 529 | 581 |
| Zero crossing | 528 | 581 |
| Derivative peak | 665 | 728 |

This substitution resulted in an insignificant change in $N_{tot}$, $(-0.2 \pm 2.5)$%. A 3.1% decrease was expected due to the lower electron density of toluene. Event and power spectra of the toluene-based scintillator were analyzed by the procedures described above. Characteristic features of spectra of the two scintillators are compared in Table I, together with light yields for $C_7H_8$ calculated relative to LS. Shifts in all features except the event peak, are consistent with a 10% increase in LY. Up-shift of the event peak, confirmed by the derivative crossing, implies a 15% increase. The conclusion here is that the position of the event peak is not a reliable measure of LY. Other spectral features give concordant values for these high LY scintillators,. A 10% loss of light is a small penalty to pay for the other more favorable properties of LAB.

Net spectra for pure LAB obtained with the LS-6500 system are shown in Fig. 4. Without added fluor or shifter, only a small fraction of the LAB emission falls in the region of PMT sensitivity. It is considered a low LY material in this study. The Compton peak in its event spectrum, N in Fig. 4(a), is reduced to a broad monotonically decreasing region above channel 10. The derivative, ND, does still identify an inflection point. However, note that $ND_{max}$ is 15% lower than $PD_{max}$. Since the positions of the two derivatives closely match for high-yield scintillators (Table I), this implies that the event derivative under estimates LYs for low-yield scintillators such as pure LAB. Comparison of Fig. 4(b) with Fig. 3(b) suggest that power spectrum, P, and its derivative, PD, approach Gaussian shapes for low LYs. Note the approximate equality of the positive and negative lobes of the PD in Fig. 4(b).

A blank has been subtracted from all spectra in this work. $N_{tot}$ and $P_{tot}$ for several types of blank spectra are compared with those for pure LAB in Table II. In the absence of real light, $N_{tot}$ and $P_{tot}$ for the black-coated vial are low. There are large increases for an uncoated vial, due to scintillations in the glass. Further increases occurred when water was added to the vial; possibly due to Compton electrons from the water exciting the glass. The event spectrum for the adopted vial+water blank is shown by open circles in Fig. 4(a). It is significant only at the lowest channels.

TABLE II: Comparison of $P_{tot}$ and $N_{tot}$ for blank light yield spectra with those for pure LAB.

| Spectrum | $N_{tot}$ | $P_{tot}$ |
|---|---|---|
| Black vial | $7.39 \times 10^2$ | $2.94 \times 10^3$ |
| Empty vial | $5.30 \times 10^4$ | $1.68 \times 10^5$ |
| Vial+$H_2O$ | $8.30 \times 10^4$ | $2.56 \times 10^5$ |
| LAB(net) | $1.04 \times 10^6$ | $1.89 \times 10^7$ |
| S/N | 12.5 | 74 |

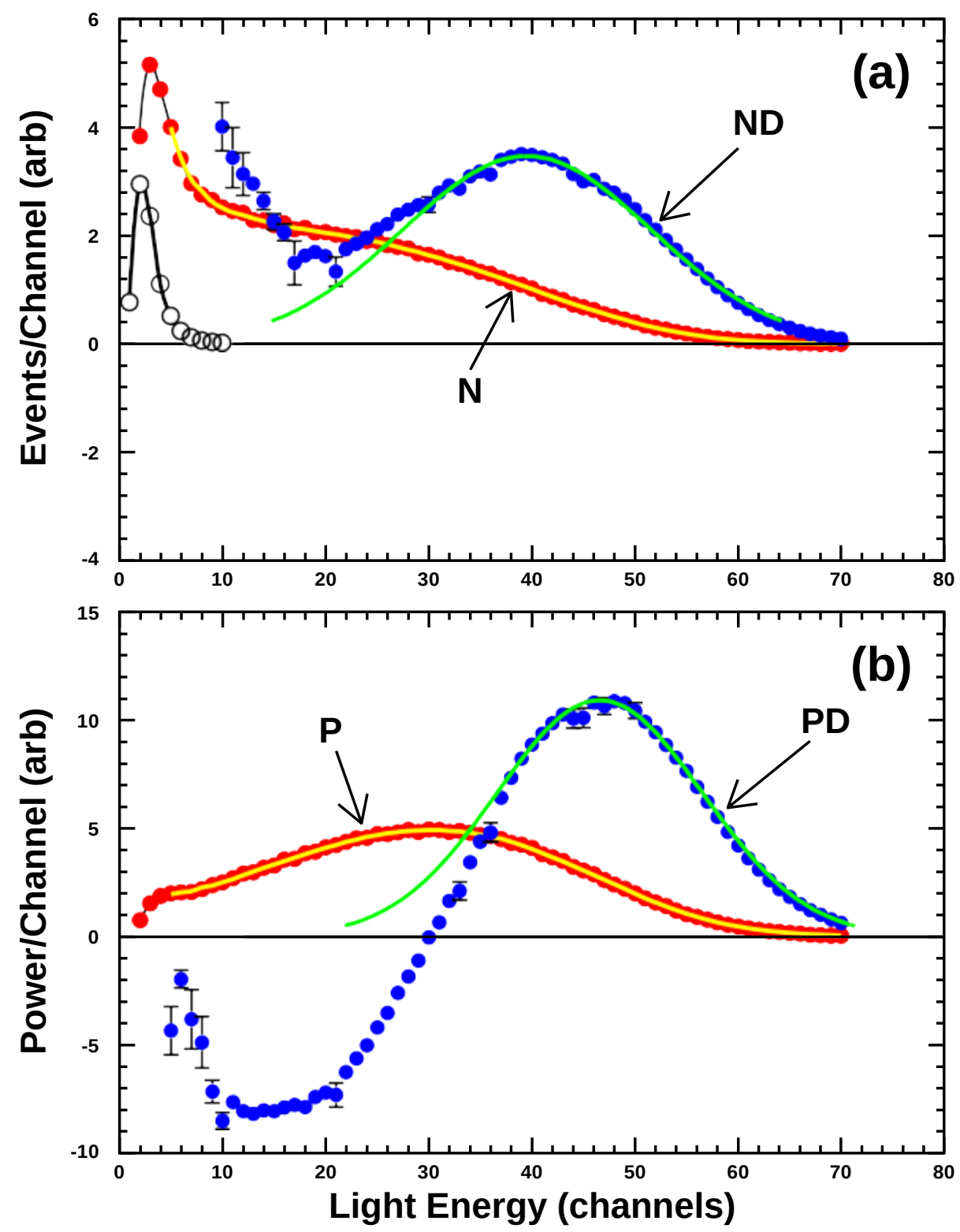

FIG. 4: (a): Light yield event spectrum for pure LAB, labeled N, and its derivative, ND. (b): The derived power spectrum, P, with its derivative,PD. An instrument blank is shown by open circles at the low-energy end of the event spectrum.

## B. PPO concentration dependence

Major emphasis of the present work focused on the binary PPO+LAB system. Values of $N_{tot}$, $P_{tot}$, $ND_{max}$ and $PD_{max}$ were obtained from 184 spectra using pro- cedures described above. These were averaged to give means at 17 PPO concentrations from 0 to 10 g/L. The number of spectra contributing to each mean varied from 5 to 29. Internal agreement of some of these data has been examined in Fig. 2.

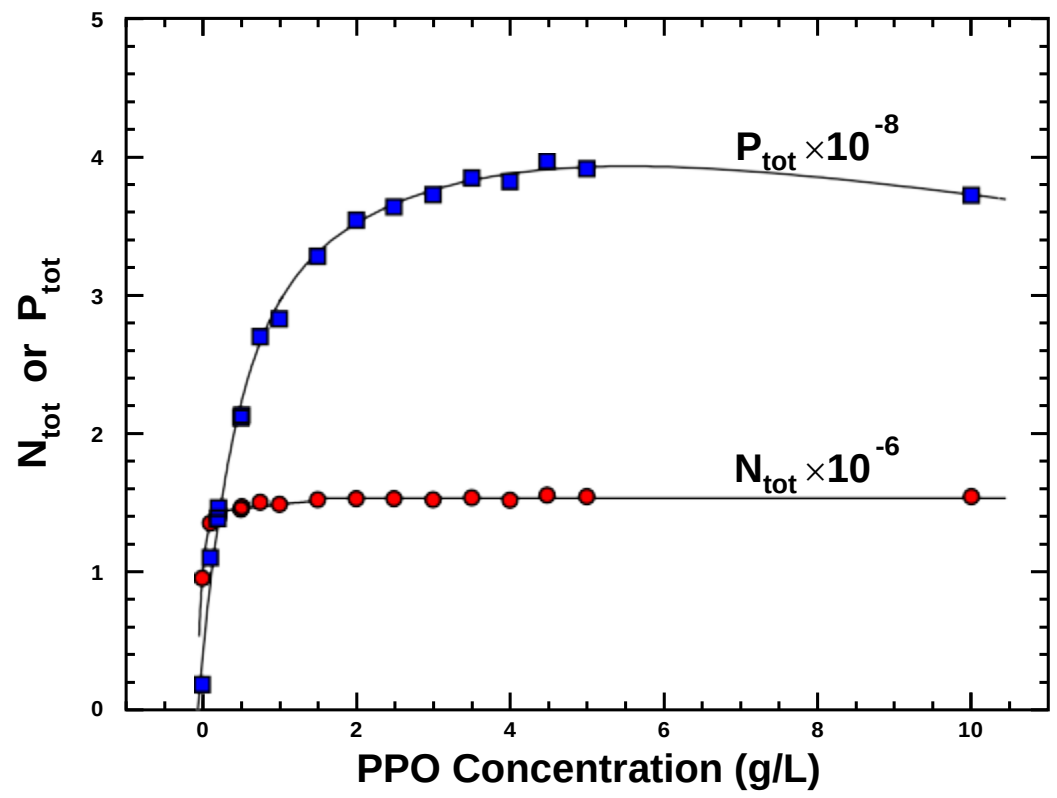

FIG. 5: Dependence of total events, $N_{tot}$, and total power, $P_{tot}$, on PPO concentration in LAB. Points are the observed values. Smooth curves show the general trends.

The dependence of $N_{tot}$ and $P_{tot}$ on PPO concentration is shown in Fig. 5. $N_{tot}$ is constant at $1.53 \times 10^6$ from 10 g/L down to 1.5 g/L. Loss of events due to the two-photon threshold increases below that concentration, reaching 38% for pure LAB. Total power rises rapidly from $1.87 \times 10^7$ for pure LAB to a maximum of $3.84 \times 10^8$ at 4.5 g/L. $P_{tot}$ is significantly lower at 10 g/L. Over the region of constant $N_{tot}$, $P_{tot}$ provides a measure of LYs without further analysis. That the maximum LY occurs at a higher PPO concentration than the 3 g/L in standard LS, confirms the conclusion of Ye et al.[6]

Light yields are shown as circles in Fig. 6 as a function of the PPO concentration. Values were obtained from $PD_{max}$, taking its peak value as the 100% reference. The general dependence of LY parallels that of $P_{tot}$ in Fig. 5. Its shape suggests that of a radioactive growth and decay chain which is described as the difference between two negative exponentials, $y = p_1 \exp(p_2 x) + p_3 \exp(p_4 x)$. As a model for light evolution, the negative (growing) component represents the absorption of LAB light by PPO, and the positive (decaying) component, the self-absorption of the PPO light. A least-squares-fit of the concentration dependent part of the LYs to that model is shown as a smooth curve in the figure. Agreement is at the 1% level when fitting concentrations from 0.1 to 10 g/L. An extrapolation of the decaying component is also shown. Differences between extrapolated values and the data points are shown as squares in the figure. They confirm the exponential absorption of LAB light by PPO. The measurements of Ye et al.[6] show that addition of MSB to the system effectively removes the decaying component by red-shifting the PPO light above its absorption spectrum.

The simple two-component model cannot account for all of the decrease of LY below 0.1 g/L. When fitting is expanded to include the point for LAB, $x^2/ndf$ increases to 6.75. The present procedure using $PD_{max}$ is well suited to examine this low-concentration region in more detail.

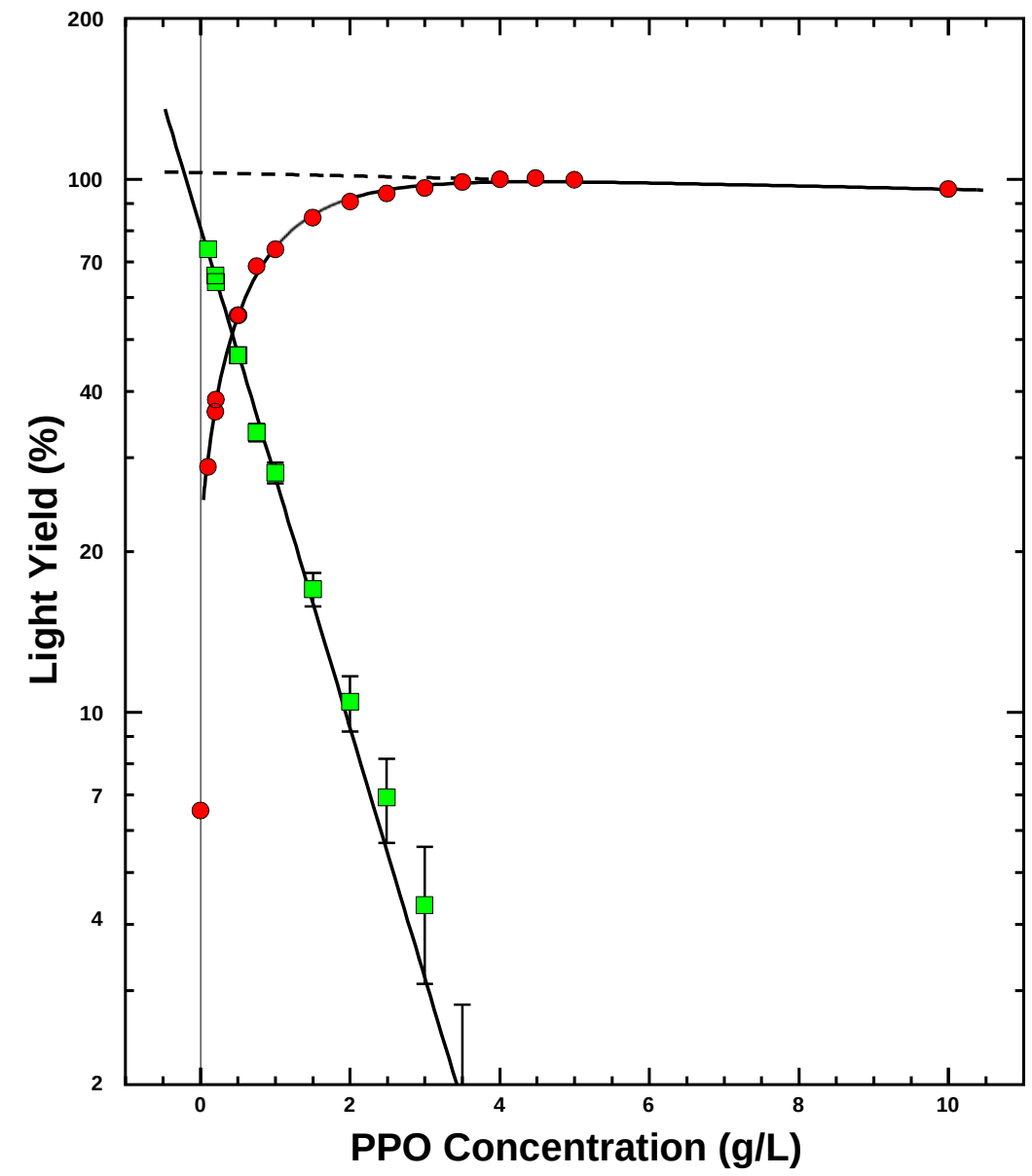

FIG. 6: Dependence of light yield on PPO concentration showing resolution into growing and decaying components. Observed values are shown by circles. The solid curve is a fit to a two-exponential model. A dashed line shows the extrapolation of the weakly-absorbing component of that model. Squares are differences between the experimental points and that extrapolation. They are consistent with the expected exponential decay.

### C. Water-based liquid scintillator

In a water-filled detector, a particle's energy and direction are obtained from Cherenkov radiation. Only particles above the Cherenkov threshold, $v/c = 1/n$ where $n$ is the refractive index, are detected. A water-based liquid scintillator (WbLS) results from adding a few percent of LS in the form of micelles to the water to allow detection of sub-threshold particles[15]. Cherenkov and scintillation light from a WbLS can be resolved on the basis of timing [16, 17]. The LS signal is “chemically” delayed from that of the prompt Cherenkov light. Light yields for two WbLS formulations have been measured using 210- and 475-MeV and 2-GeV protons from the AGS.[18] We thought it of interest to see whether the much less complicated power-derivative procedure described above could be applied to WbLS.

Values of $P_{tot}$ for the PPO+LAB system are plotted as small points in Fig. 7 as function of LY determined from $PD_{max}$. The solid line is a fit to $y = p_1 x$ for points with LY> 25%. The point for pure LAB, LY=6.5%, falls 26% below that line, suggesting that its power spectrum is cutoff by the two-photon threshold. Results for two WbLS formulations (WbLS-1 and WbLS-5) are shown as larger points in the figure. The event and power spectra of WbLS-5 are similar to those of pure LAB. Its point at LY= 5.7% is 21% below the line. The event spectrum of WbLS-1 shows extreme effects of resolution. There is no derivative peak, only a shoulder. The point at LY= 2.3% in Fig. 7 is based on the power spectrum. It falls 29% below the line.

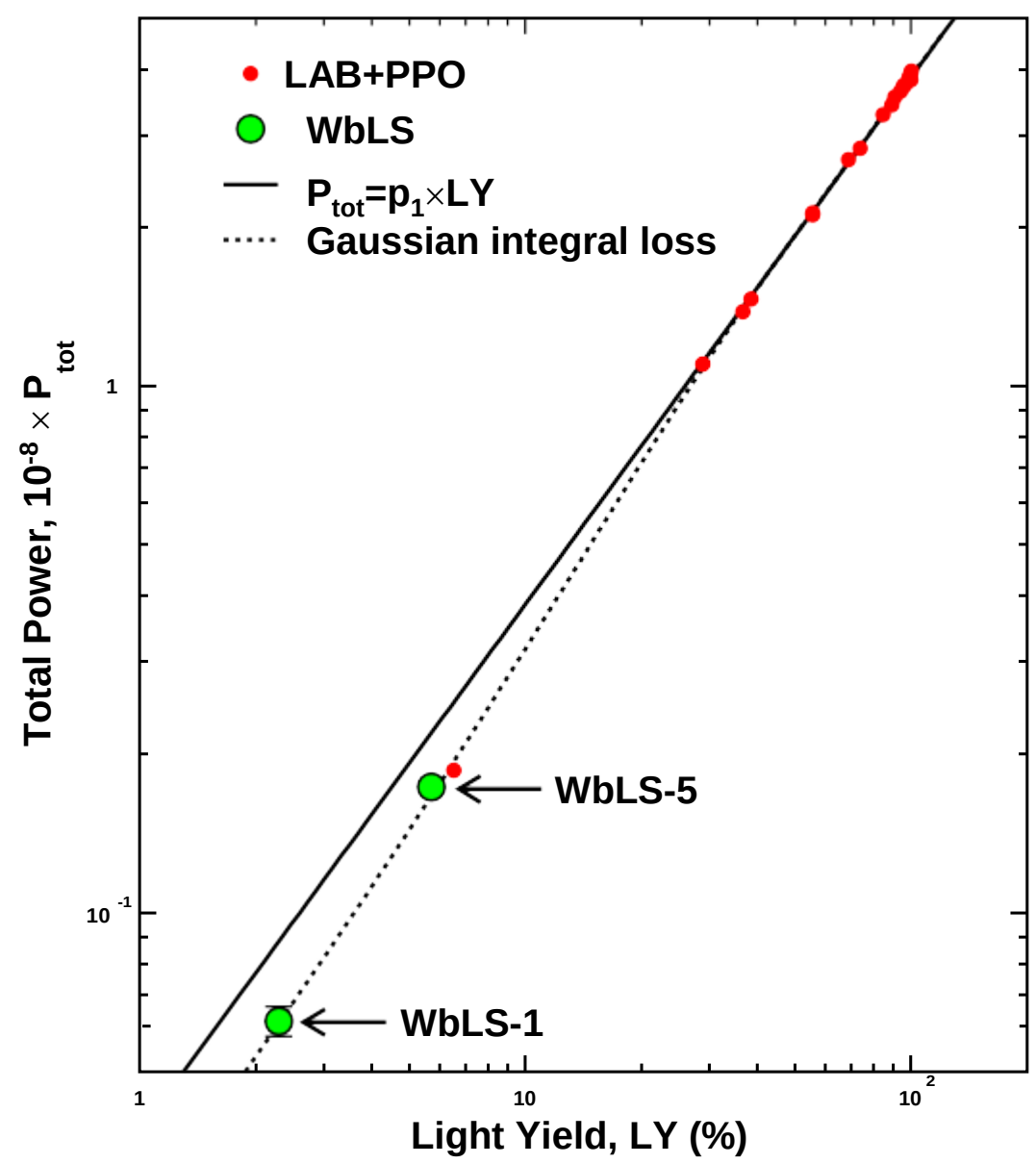

FIG. 7: Total power, $P_{tot}$, as a function of light yield for LAB+PPO (small points) and water-based liquid scintillator (large points). The solid line showing proportionality between $P_{tot}$ and LY, is based on points with LY> 25%. The lowest LAB+PPO point and two for WbLS fall below that line. The dashed curve shows the behavior expected if $P_{tot}$ had a Gaussian integral loss, see text.

The dashed curve in the figure includes an approximate correction for losses of $P_{tot}$ due to the two-photon threshold. Were a power spectrum Gaussian and the cutoff sharp, the loss of power would be described by a Gaussian integral. It would be zero for a spectrum located well above the threshold, but rise to 50% for one centered on it. The dashed line in Fig. 7 shows the prediction of that model. Since losses are from the low end of a power spectrum and $PD_{max}$ is determined from the upper end, it is reasonable to assume a LY determined from $PD_{max}$ even as low as that for WbLS-1 would be reliable, albeit with large errors.

## V. CONCLUSIONS

Combination of the power derivative procedure with the LS-6500 provides a powerful tool for routine light yield determinations in a laboratory setting. Values down to 2% of those for high-yield scintillators can be determined using small samples. Results for LAB+PPO based scintillators show consistency with a simple model for light evolution in that system. They do suggest additional studies to resolve the behavior at PPO concentrations < 0.1 g/L. The procedure should be of value during further development of the WbLS.

### Acknowledgments

This work, conducted at Brookhaven National Laboratory, was supported by the U.S. Department of Energy (DOE) under Contract No. DE-SC0012704.